%% file: ms.tex
\documentclass{emulateapj}

\newcommand{\co}{$^{12}$CO~}
\newcommand{\tco}{$^{13}$CO~}

\shorttitle{Embedded Young Stellar Object Candidates in W51}
\shortauthors{Kang et al.}

\begin{document}

\title{Embedded Young Stellar Object Candidates in the Active Star
       Forming Complex W51: Mass Function and Spatial Distribution}
\slugcomment{Accepted for publication in ApJ}
\author{\sc Miju Kang\altaffilmark{1,2,3},
            John H. Bieging\altaffilmark{3},
            Matthew S. Povich\altaffilmark{4,5},
            and Youngung Lee\altaffilmark{1}}
\altaffiltext{1}{International Center for Astrophysics,
                 Korea Astronomy and Space Science Institute,
                 Hwaam 61-1, Yuseong, Daejeon 305-348, South Korea;
                 mjkang@kasi.re.kr}
\altaffiltext{2}{Department of Astronomy and Space Science,
                 Chungnam National University, Daejeon 305-764, South Korea}
\altaffiltext{3}{Steward Observatory,
                 University of Arizona,
                 933 North Cherry Avenue, Tucson, AZ 85721}
\altaffiltext{4}{Department of Astronomy, University of Wisconsin at Madison,
                 475 N. Charter Street, Madison, WI 53706}
\altaffiltext{5}{NSF Astronomy and Astrophysics Postdoctoral Fellow,
                 Department of Astronomy and Astrophysics, Pennsylvania State
                 University, 525 Davey Lab, University Park, PA 16802}

\begin{abstract} 

We present 737 candidate Young Stellar Objects (YSOs) near the W51 Giant
Molecular Cloud (GMC) over an area of $1.25 \arcdeg \times 1.00\arcdeg
$ selected from {\it Spitzer Space Telescope} data. We use spectral
energy distribution (SED) fits to identify YSOs and distinguish them
from main-sequence or red giant stars, asymptotic giant branch stars,
and background galaxies. Based on extinction of each YSO, we separate
a total of 437 YSOs associated with the W51 region from the possible
foreground sources. We identify 69 highly embedded Stage 0/I candidate
YSOs in our field with masses ${>}~5 ~M_{\odot}$ (corresponding to mid-
to early-B main-sequence spectral types) 46 of which are located in the
central active star forming regions of W51A and W51B. From the YSOs
associated with W51, we find evidence for mass segregation showing
that the most massive YSOs are concentrated on the W51 \ion{H}{2}
region complex. We find a variation in the spatial distribution of the
mass function (MF) of YSOs in the mass range between 5 $M_\odot$ and
18 $M_\odot$. The derived slopes of the MF are $-1.26$ and $-2.36$ in
the active star-forming region and the outer region, respectively. The
variation of the MF for YSOs embedded in the molecular cloud implies
that the distribution of stellar masses in clusters depends on the local
conditions in the parent molecular cloud.

\end{abstract}

          
%%%%%%%%%%%%%%%%%%%%%%%%%%%%%%%%%%%%%%%%%%%%%%%%%%%%%%%%%%%%%%%%%%%%%%%%%%%
\section{Introduction}

Stars form in various environments. High-mass stars are especially
important because they affect their environment through such phenomena as
outflows, stellar winds, and supernovae. To understand the interaction
of stars and the interstellar medium (ISM) it is necessary to
identify embedded young stellar objects (YSOs) in the parent molecular
cloud. Therefore taking a complete census of YSOs is an important step
toward understanding YSOs and their environment on a large scale.

W51 is an active star-forming region located in the Sagittarius
spiral arm. The distance is uncertain, with published values including
$5.1^{+2.9}_{-1.4}$ kpc \citep{Xu09}, $6.1\pm1.3$ kpc \citep{Imai02},
or $8.3\pm2.5$ kpc \citep{Schneps81}. The radio continuum sources
comprising W51 are within a massive giant molecular cloud (GMC)
\citep{Mufson79}. Observations showing the star forming activity
have been carried out in various wavelengths of radio continuum, HI
line, molecular lines, water masers, near-IR, X-ray \citep{Bieging75,
Carpenter98, Imai02, Koo97a, Koo97, Koo99, Mehringer94, Okumura00}. W51
is a region of ongoing massive star formation \citep{Lacy07, Zapata08} and
therefore a good place to study the feedback between stars and the ISM.

To understand the interaction of molecular clouds and YSOs in the W51
\ion{H}{2} region complex we mapped a 1.25\arcdeg\ $\times$ 1.0\arcdeg\
region in the $J=2-1$ transitions of \co and \tco with the 10 m Heinrich
Hertz Telescope (HHT) on Mount Graham, Arizona. The molecular line maps
will be presented in a separate paper (Bieging et al., in preparation). In
the present paper, we use {\it Spitzer} data for identifying the
YSO candidates associated with the molecular clouds. {\it Spitzer}
surveys in the mid-IR allow sensitive studies of deeply embedded star
formation. Although extinction is significantly reduced in the mid-IR,
it is still important for active star forming regions in giant molecular
clouds. In particular, the W51 region shows no detectable increase in
star counts because of severe extinction \citep{Benjamin05}. Therefore
we have to consider extinction effects carefully in identifying and
characterizing candidate YSOs. Combining the YSOs selected from the {\it
Spitzer} photometry and molecular clouds with kinematic information, we
can examine the feedback process between star(s) and ISM. The detailed
analysis for the interaction of YSOs and associated molecular clouds
will be presented in other publications (\cite{Kang09}; Kang et al.,
in preparation). In this paper, we tabulate all of the YSO candidates
and discuss their spatial distribution and mass function. In Section
\ref{sec:yso_selection}, we introduce the data sets used and describe
our methods for selecting and classifying candidate YSOs. We discuss
some of the implications of our results in Section \ref{sec:result}
and summarize the results in Section \ref{sec:summary}.

%%%%%%%%%%%%%%%%%%%%%%%%%%%%%%%%%%%%%%%%%%%%%%%%%%%%%%%%%%%%%%%%%%%%%%%%%%%
\section{YSO Selection and Classification} 
\label{sec:yso_selection}

\subsection{Data} 

The Galactic Legacy Infrared Mid-Plane Survey Extraordinaire \cite[GLIMPSE
I;][]{Benjamin03} survey covered the Galactic plane ($ 10 \arcdeg <|l|<
65 \arcdeg,\, |b| < 1\arcdeg$) with the four mid-IR bands (3.6, 4.5, 5.8,
and 8.0 \micron) of the Infrared Array Camera \cite[IRAC;][]{Fazio04} on
the Spitzer Space Telescope. Each IRAC band contains different spectral
features : Band 1 $[3.6]$ shows mainly continuum emission from stars, Band
2 $[4.5]$ traces H$_{2}$ rotational transitions arising in the shocked gas
associated with outflows, and emission in Bands 3 $[5.8]$ and 4 $[8.0]$
is dominated by polycyclic aromatic hydrocarbon (PAH) features. For this
study, we have retrieved images of the 1.25\arcdeg\ $\times$ 1.0\arcdeg\
region of W51 centered on ($l,b$) = ($49.375 \arcdeg,\, -0.2 \arcdeg$)
(note that this is the region we mapped in CO) by combining two mosaic
images which have a high resolution (1\farcs2 pixel$^{-1}$) in all four
IRAC bands (Figure \ref{All_W51}). For this same region, we used the
GLIMPSE I Catalog including the sources detected at least twice in one
band with a S/N $\ge$ 5 and at least once with a S/N $>$ 3 in an adjacent
band. Generally, the flux density limits insure that the detections are
$>$ 5$\sigma$. The 5$\sigma$ point-source detection limits of the GLIMPSE
are nominally 0.2, 0.2, 0.4, and 0.4 mJy for the IRAC 3.6, 4.5, 5.8,
and 8.0 \micron\ bands, respectively. These limits are significantly
higher in regions of bright diffuse emission. The GLIMPSE I Catalog also
tabulates $JHK_{s}$ flux densities from the 2MASS point source catalog
\citep{Skrutskie06} for all GLIMPSE sources with 2MASS identifications.

%------------------------------
\input{fig01.tex}

%------------------------------

MIPSGAL \citep{Carey05} is a legacy program covering the inner Galactic
plane, $10\arcdeg < |l| < 65\arcdeg$ for $ |b| < 1\arcdeg$, at 24 and 70
\micron\ with the Multiband Imaging Photometer for Spitzer Space Telescope
\cite[MIPS;][]{Rieke04}. The resolution of the 24 \micron\ mosaics from
the MIPSGAL survey is 2\farcs4 pixel$^{-1}$. We extracted 24 \micron\
point-sources with $F/{\delta}F >$ 7, then bandmerged the 24 \micron\
sources with the GLIMPSE Catalog sources using a 2\farcs0 correlation
radius. The final source list consists of all 8 bands combined: 2MASS,
GLIMPSE, and MIPSGAL 24 \micron. A total of 104,582 sources within the
$1.25\arcdeg \times 1.0 \arcdeg$ region of W51 centered on ($l,b$)
= ($49.375 \arcdeg,\,-0.2 \arcdeg$) were selected from the GLIMPSE
I Catalog.

%---------------------------------------------------------------------------
\subsection{Color Selection} 

The colors of most main sequence (MS) stars are near zero for any
combination of the IRAC and MIPS bands, while YSOs have red colors due
to the emission from surrounding warm dust. As an initial approach,
we applied a well defined color selection criteria of \cite{Simon07}
to our source list for identifying YSO candidates associated with the
molecular cloud around W51. After selecting candidate YSOs from the region
in color-color space defined by \cite{Simon07}, we classified them with
the spectral index $\alpha$ in the IRAC bands. The spectral index is
\begin{equation} 
\alpha = \frac{d\ \mathrm{log}\ [\lambda F({\lambda})]}
              {d\ \mathrm{log} {\lambda}},
\end{equation} 
where $\lambda$ is the wavelength and $F(\lambda)$ is the flux at that
wavelength. We obtain $\alpha$ from a linear fit to the logarithm of the
fluxes for all 4 IRAC bands and then classify YSOs as Class I sources with
$0.3 \leq \alpha_{IRAC}$, `` Flat '' sources with $-0.3 \leq \alpha_{IRAC}
< 0.3$, Class II sources with $-1.6 \leq \alpha_{IRAC} < -0.3$, and Class
III sources with $\alpha_{IRAC} < -1.6$ \citep{Lada87,Greene94}.

%------------------------------
\input{fig02.tex}

%------------------------------

%------------------------------
\input{table01.tex}

%------------------------------

Figure \ref{ccd_f2}({\it a}) shows the distribution of YSOs within the
color selection criteria of \cite{Simon07} represented by the solid
line on the IRAC $[3.6]-[4.5]$ vs. $[4.5]-[8.0]$ color-color diagram. In
Figure \ref{ccd_f2}({\it a}), a large number of YSO candidates with red
colors were not selected because the \cite{Simon07} color criteria were
designed to identify a relatively clean sample of YSOs. Many reddened main
sequence stars in our W51 field were identified as YSOs, however, because
the \cite{Simon07} color criteria were derived from the Small Magellanic
Cloud which is not affected by the large interstellar extinction of the
Galactic plane. The W51 region is several kpc from the Sun in the Galactic
plane so interstellar extinction is significant. Figure \ref{ccd_f2}({\it
b}) shows the IRAC $[3.6]-[4.5]$ vs. $[5.8]-[8.0]$ color-color diagram
by \cite{Simon07} with YSO categories based on $\alpha$ for comparing
with the distribution of YSOs found by other selection criteria in the
same color-color space.

Figure \ref{ccd_f2}({\it c}) shows the distribution of YSOs identified
with the IRAC color criteria of \cite{Gutermuth08}. They classified
the YSO candidates into two groups after removing sources dominated by
PAH emission, shock emission and broad-line AGNs. One group is Class I
protostars and the other is the more evolved Class II sources. Applying
color classification of \cite{Gutermuth08}, we found that the YSOs
candidates in the W51 region separate into 366 Class I sources and 1457
Class II sources. Although these color criteria selected the interesting
YSO candidates having very red colors, the sample also contained a large
number of reddened stellar photospheres because the color criteria were
derived from the study of the nearby star forming region, NGC 1333.

Class I, Flat, and Class II sources in Figure \ref{ccd_f2}({\it b}) belong
to Class I and relatively red Class II samples in Figure \ref{ccd_f2}({\it
c}). Color selection criteria may be suitable as a method to select
the objects in a relatively young stage. However, more evolved sources,
which include the Class III objects in Figure \ref{ccd_f2}({\it b}) and
those Class II objects around zero color in Figure \ref{ccd_f2}({\it c}),
may be mis-identified because of the large interstellar extinction toward
the W51 region.

%---------------------------------------------------------------------------
\subsection{SED fitting}   

%------------------------------
\input{table02.tex}

%------------------------------
%------------------------------
\input{table03.tex}
%------------------------------
%------------------------------
\input{table04.tex}

%------------------------------

We used the spectral energy distribution (SED) fitting tools of
\cite{Robitaille07} for identifying and classifying YSOs in the W51
giant molecular cloud complex. The SED fitting tool works as a regression
method to find the SEDs within a specified $\chi^{2}$ from a large grid
of models after fitting the input data points. The grid of models contains
stellar atmospheres, a limited set of spectra of galaxies, AGB stars, and
YSO models. The grid of YSO models was computed by \cite{Robitaille06}
using the 20,000 2-D radiation transfer models from \cite{Whitney03a,
Whitney03b,Whitney04}. Each YSO model has SEDs for 10 viewing angles
(inclinations), so the total YSO grid consists of 200,000 SEDs. The SED
fitting tool provides the evolutionary stage and physical parameters
such as disk mass, disk accretion rate, and stellar temperature of
YSOs. The SED fitting tool has been used to study star forming regions
in various environments, e.g., M17 \citep{Povich09}, the Eagle Nebula
\citep{Indebetouw07}, IRAS 18507+0121 \citep{Shepherd07}, and the Small
Magellanic Cloud \citep{Simon07}.

Table \ref{table1} summarizes the number of sources. We reset the
photometric uncertainties to a value of ${\delta}F/F = 10~\%$ for all
GLIMPSE Catalog sources, and to 15 \% for the MIPS 24 $\micron$ fluxes
including systematic errors based on the GLIMPSE documents\footnote{
http://irsa.ipac.caltech.edu/data/SPITZER/GLIMPSE/doc/}. These values
were adopted instead of the formal errors in the Catalog in order to fit
without possible bias caused by underestimating the flux uncertainties. A
total of 70,767 sources for SED fitting was detected in $N_{data} \geq 4$
of the 8 IR bands in the GLIMPSE Catalog combined with MIPS 24 $\micron$
fluxes. Our final goal is to find YSOs associated with molecular clouds,
so we start by removing stars. The foreground extinction, up to an
allowed maximum $A_{V}$ of 30 mag in this study is accounted for with
an extinction law derived from GLIMPSE data \citep{Indebetouw05} in the
process of fitting stellar photosphere SEDs. We calculate a best-fit
$\chi^{2}$ normalized by the number of data points between 4 and 8 used
in the fit. We consider all sources with $\chi^{2}/N_{data} \leq 4$ as
good fits to reddened stellar photospheres. The total of sources with
$N_{data} \geq 4$ separates into 69,316 well fit sources and 1,451 badly
fit sources by this threshold value. In most cases, even highly reddened
stars are classified as good fits because the fitting tool includes the
extinction law.

Next, we fit 1,451 sources that are poorly fit by reddened stellar
photospheres using the YSO models of \cite{Robitaille06}. We allow the
distance range to be from 5 to 9 kpc and the interstellar extinction from
0 to 60 mag in $V$ band for fitting YSOs. In the same manner as fitting
stellar photospheres, we consider sources with $\chi^{2}/N_{data} \leq 4$
to be well-fit. This YSO fitting process separates these 1451 sources into
1327 good fit YSO sources and 124 bad fit sources. Among good fit sources
we discard those not detected at 24 $\micron$ but which show 8 $\micron$
flux excesses above an extrapolation of the 3 shorter wavelength IRAC
bands. These sources with 8 $\micron$ excess emission are usually found
to be a noise peak or a diffuse background feature. In cases which show
an IR excess in the MIPS 24 $\micron$ band only, we also discard the
source, because these may be a false match due to the difference in
resolution between GLIMPSE and MIPSGAL. We extract 14 bright sources
which may be evolved stars on the asymptotic giant branch (AGB). For
example, carbon stars enshrouded by dusty envelopes have SEDs similar to
YSOs. Most of the 124 objects classified as bad fits ($\chi^{2}/N_{data}
\geq 4$) are stars with saturated fluxes or signs of variability or bad
photometry. After replacing the uncertainties of sources with questionable
fluxes to upper or lower limits, we moved five YSO candidates among bad
fit sources to the ``well-fit'' sample. Two samples were moved to the
AGB candidate category. After all these adjustments, we distinguish 737
good YSO candidates, 576 bad samples, and 16 bright sources which are
likely to be AGB stars.

We divide the final 737 YSO candidates into three evolutionary stages
defined by the well-fit model disk mass $M_\mathrm{disk}$ and the
envelope accretion rate $\dot{M}_\mathrm{env}$, both normalized
by the mass of the central star $M_{\star}$ ;  Stage 0/I with
$\dot{M}_\mathrm{env}/M_{\star} > 10^{-6}~\mathrm{yr}^{-1}$, Stage II
YSOs with $\dot{M}_\mathrm{env}/M_{\star} < 10^{-6}~\mathrm{yr}^{-1}$
and $M_\mathrm{disk}/M_{\star} > 10^{-6}$, and Stage III with
$\dot{M}_\mathrm{env}/M_{\star} < 10^{-6}~\mathrm{yr}^{-1}$ and
$M_\mathrm{disk}/M_{\star} < 10^{-6}$. Stage 0/I, II, and III YSOs have
significant infalling envelopes, optically thick disks, and optically thin
disks, respectively \citep{Robitaille06}. We determine the evolutionary
stage of each source using the relative probability distribution for
the stages of all the ``well-fit'' models. The well-fit models of each
source are defined by \begin{equation} \chi^2 - \chi^2_{\mathrm{min}} \leq
2N_{\mathrm{data}}, \end{equation} where $\chi^2_{\mathrm{min}}$ is the
goodness-of-fit parameter for the best-fit model. The relative probability
of each well-fit model is estimated according to \begin{equation}
P(\chi^2)=e^{-(\chi^2-\chi^2_{\mathrm{min}})/2} \end{equation} and is
normalized. After a probability distribution for the evolutionary stage
of each source is constructed from the Stages of all the well-fit models,
the most probable stage of each source is determined by the condition
that $\Sigma P(\mathrm{Stage}) \geq 0.67$. If this condition is not
satisfied, then the stage of the source is classified as ambiguous,
though the object is still counted as an YSO.

From well-fit models for each source derived from the SED fitting tool,
we calculated the $\chi^2$-weighted model parameters: interstellar
extinction to the source ($A_V$), stellar mass ($M_\star$), and
total luminosity ($L_{tot}$ : total bolometric luminosity, including
luminosity produced by accretion). Note that $A_V$ is the foreground
interstellar extinction, which does not include the extinction produced
by a circumstellar disk or envelope. The $\chi^2$-weighted average
$A_V$ distribution for all YSO candidates shows the linear variation
in the direction of increasing extinction. We consider 437 YSOs with
more than 10 mag $A_{V}$ of foreground extinction as sources associated
with the W51 complex because the foreground extinction toward the W51
complex is $A_{V} \sim 20$ mag \citep{Goldader94}. The majority (97\%
(192/198)) of the YSO candidates detected in all 2MASS bands have $A_V$
values $< 10$ mag, and are classified as foreground objects;  while 95\%
(295/312) of the YSO candidates detected only in IRAC/MIPS bands have
fitted $A_V \ge 10$ mag and are classified as associated with the W51
region. We list magnitudes and classifications for the entire list of
737 YSO candidates in Table \ref{table:mag} . Foreground sources and W51
sources are denoted as ``F'' and ``W'', respectively. We present some of
the derived physical parameters from the SED fitting for all 737 YSOs in
Table \ref{table:parameters}.  We report lower and upper ranges of 68\%
probability for each candidate YSO. In cases where the range in the 68\%
probability distribution is very small, we list ranges for the 95\%
confidence intervals and mark these with an asterisk.

In the last three columns of Table \ref{table:parameters}, we give
evolutionary stages and SED classes for each YSO candidates. We classify
the 737 YSO candidates as 228 Stage 0/I, 255 Stage II, 5 Stage III, and
249 ambiguous YSO candidates based on the disk mass $M_\mathrm{disk}$ and
the envelope accretion rate $\dot{M}_\mathrm{env}$ of each source. The
spectral index $\alpha_{IRAC}$ is the value derived from the flux
densities of the sources detected in all four IRAC bands. Among 737 YSO
candidates 561 sources are detected in all four IRAC bands and divided
into 83 Class I, 164 Flat, 195 Class II, and 19 Class III. Using the flux
densities available between 2 and 24 \micron\ to derive the spectral
index, $\alpha_{2-24}$, all sources are classified as 256 Class I,
243 Flat, and 235 Class II, and 3 Class III. We summarize the physical
Stages and observational Classes of YSOs in Table \ref{table:stage_class}.

%------------------------------
\input{fig03.tex}

%------------------------------
Figure \ref{ccd_f2}({\it d}) shows the distribution of 561 YSO candidates
detected in all 4 IRAC bands in the IRAC $[3.6]-[4.5]$ vs.$ [5.8]-[8.0]$
color-color diagram for comparison with Figure \ref{ccd_f2}({\it b}) and
({\it c}). A total of 176 YSO candidates (24\% of the total YSOs) are
identified from a combination of JHK$_\mathrm{s}$ and MIPS 24 \micron\
fluxes although these are not detected in all 4 IRAC bands. In Figure
\ref{ccd_f2}({\it d}), YSO candidates are classified according to their
most probable evolutionary Stage: red for Stage 0/I, yellow for Stage
II, and cyan for Stage III. Since Stage III YSOs with optically thin
remnant disks and their SEDs are dominated by photospheric emission,
those objects are difficult to identify by IR excess alone. Despite a low
probability, 5 Stage III sources are in our final samples because the
evolutionary Stage is determined from multiple YSO models of each YSO
candidate statistically. Green dots are for ambiguous sources that are
YSO candidates but do not have a sufficiently well-defined probability
distribution to determine the evolutionary Stage. Blue dots are sources
that were well-fit by YSO SEDs but were not included in the final
sample. Figure \ref{contamination}, {\it top} shows examples of these
rejected sources with IR excess emission only in the IRAC 8.0 \micron\
band ({\it{left}}) and with contamination by diffuse background emission
({\it{right}}). Examples of SEDs for two AGB candidates are plotted in
Figure \ref{contamination}, {\it bottom}.

%---------------------------------------------------------------------------
\subsection{Estimating Contamination}
\label{sec:est_cont}

Because other IR-excess populations (e.g., AGB stars and galaxies)
have colors or SEDs very similar to YSOs, the SED-fitting method will
inevitably have contamination from these sources. Therefore we use an
empirical method to estimate the contamination of YSO candidates from
galaxies and dusty evolved AGB stars. GLIMPSE is a shallow survey of
the Galactic plane while c2d is a deep survey of nearby star forming
regions away from the Galactic plane: Chameleon II, Lupus, Perseus,
Serpens, and Ophiuchus \citep{Evans09}. The contamination by external
galaxies is serious in deep IRAC observations, such as the c2d Survey
\citep{Harvey07,Porras07}. In GLIMPSE, evolved stars like AGB stars with
dusty winds and unresolved planetary nebulae (PNe) may be more serious
contamination sources than external galaxies \citep{Povich09}.

%------------------------------
\input{fig04.tex}

%------------------------------

To estimate an upper limit for evolved star (AGB) contamination,
we use the color criterion ($ [8.0] - [24] < 2.2~\mathrm{mag} $) of
\cite{Whitney08}. Their color-magnitude criterion derived from the
LMC is not appropriate for W51, given the difference in distance and
the fact that the AGB stars will not all be at the same distance as the
YSOs. Note that relatively few extreme AGB stars are expected to populate
the upper-right region of the $[8.0]$ vs. $[8.0]-[24]$ color-magnitude
diagram, and in the Galaxy such sources are likely to be saturated in
GLIMPSE. Sources with $ [8.0] - [24] < 2.2~\mathrm{mag} $ are 10 of our
345 YSO candidates detected in [8.0] and [24] \micron\ bands, or $\sim 3
\%$ (Figure \ref{harvey_cont}($b$)). They do yield a similar (low) surface
density of AGB stars as found for the M17 region by \cite{Povich09}. The
distribution of red sources in the Galactic midplane \citep{Robitaille08},
which consist mostly of YSOs and AGB stars, also shows that the number
of YSO candidates represents a significant enhancement toward the
W51 region while the surface density of candidate AGBs is very low in
our region of $1.25\arcdeg \times 1.00 \arcdeg$ centered on ($l,b$) =
(49.375\arcdeg, $-$0.2\arcdeg).

For each source detected in all IRAC bands and in MIPS 24 \micron, we
calculated the probability of being an extragalactic contaminant using the
formulas in \cite{Harvey07}. These formulas were derived for the Serpens
molecular cloud field, which has an average extinction of $A_V \sim 7.5$
mag, from the distribution found by \cite{Harvey07}.  The Serpens cloud
lies at $b = 5$\arcdeg\ latitude, so almost all of the extinction must
be associated with the cloud. In contrast, our W51 field is right in the
plane of the Galaxy so the line of sight to any external galaxy should
have a minimum extinction corresponding to at least 25 kpc in the plane.
We assume a minimum of 1 mag/kpc or $A_V \geqq 25$ mag for our field or
17.5 mag more than used for the Serpens field.  We therefore adjust the
\cite{Harvey07} criteria by the expected additional extinction at [4.5],
[8.0], and [24], using the extinction laws of \cite{Indebetouw05} and
\cite{Flaherty07}. These give an additional 0.7 mag in all three of the
IR bands. We add this amount to the constants for the 3 IR magnitudes
in the formulae of \cite{Harvey07}. When we applied this modified set
of color-magnitude criteria to 305 YSO candidates detected in all IRAC
bands and in MIPS 24 \micron, 35 sources were in the galaxy-candidate region
of the probability histogram using the \cite{Harvey07} cutoff of log
P = $-1.47$.  We note, however, that this cutoff represents a {\it very}
conservative estimator, i.e., a probability as low as 0.034 counts as a
galaxy candidate.  Figure \ref{harvey_cont} presents three color-magnitude
diagrams showing the possibility of contamination using the modified
\cite{Harvey07} criteria, where the solid lines from their formulae are
shifted down by 0.7 mag in the magnitude axes. The joint probability
histogram in Figure \ref{harvey_cont}($d$) shows that, of the 35 sources
in the nominal GALc region, only 8 occupy the bins with a probability
greater than 0.06 (log P $> -1.25$) of being an extragalactic contaminant,
i.e., about 3\% of the sample of 305 objects.

We therefore estimate an {\it upper limit} of about 6\% to the
contamination by galaxies and AGB stars in our list of YSOs. This limit
implies a smaller contamination fraction compared to other regions,
e.g., 10\% for M17 \citep{Povich09}, 40\% for the intrinsically red
sources in the Galactic midplane \citep{Robitaille08}, and 55\% for the
LMC \citep{Whitney08}. Therefore, considering up to 2\% contamination by
unresolved Planetary Nebulae \citep{Robitaille08,Whitney08}, we estimate
an {\it upper limit} to the contamination of 8\% in our YSO candidates.

%%%%%%%%%%%%%%%%%%%%%%%%%%%%%%%%%%%%%%%%%%%%%%%%%%%%%%%%%%%%%%%%%%%%%%%%%%%
\section{Results and Discussion}
\label{sec:result}

Our target area contains very active star-forming regions in the disk
of the Milky Way (Figure \ref{All_W51}). We find a total of 737 YSO
candidates within the region around the W51 complex based on the SED
fitter. After dividing 737 YSOs into foreground and W51 YSOs based on
the amount of foreground extinction, we focus on 437 YSO candidates
associated with the W51 complex. 

These 437 sources divide into 199 Class I, 168 Flat, and 70 Class II when
classified by the empirical SED spectral index, $\alpha_{2-24}$. For
comparison, the percentage of sources in these observational SED
classes for all c2d clouds is reported as 16\% Class I, 12\% Flat,
60\% Class II, and 12\% Class III \citep{Evans09}. The ratio of Class
I plus Flat spectrum to Class II sources in W51 is about 5, which is
completely different from the low mass star-forming regions such as Cha
II (0.16), Lupus (0.29), Perseus (0.57), Serpens (0.42), and Ophiuchus
(0.47) \citep{Evans09}. The physical ``Stages'', I, II, and III should
roughly correspond to the observational ``Classes'' I+Flat, II, and III
\citep{Harvey07}. In W51 the 168 Flat sources divide into 30 Stage 0/I,
75 Stage II, and 63 with ambiguous stage. Using the physical evolutionary
models for all sources associated with W51, 128 are classified as
Stage 0/I, 162 as Stage II, and 147 as ambiguous sources. The massive
star-forming region W51 appears to contain YSOs in a very early stage as
indicated by both observational and physical classifications. The ratio
of Stage 0/I to II for the entire set of YSOs in the large $1.25\arcdeg
\times 1.00\arcdeg$ region is similar to that for other massive star
forming regions like M17 \cite[25/40;][]{Povich09}, N66 in the SMC
\cite[33/50;][]{Simon07}, and the LMC \cite[145/147;][]{Whitney08}.

%------------------------------
\input{fig05.tex}

%------------------------------

In the central active star-forming region (inside the boxed area in
Figure \ref{w51_mass_plot}) the ratio of Stage 0/I to II for all YSOs
is 2.56 (64/25). In the same region this ratio for the massive YSOs
($> 5M_{\odot}$) only is 2.71 (46/17). The ratio of YSOs outside the
central bright region is 0.47 (64/137) and the ratio for sources greater
than $5M_{\odot}$ in the same region is 0.92 (23/25).  The increase
in Stage 0/I to II is due in part to a bias toward detecting sources
with positive spectral indices in regions of bright mid-IR nebular
emission. M17 provides an extreme example of bright mid-IR nebular
emission, where the ratio reaches 1.3 (13/10) in the central bright
region \citep{Povich09}.  In both W51 and M17, the relative fraction
of sources with Ambiguous Stage determinations increases in the bright
nebular regions, due to the decreased sensitivity of the MIPS 24 \micron\
point-source detections, an important discriminant between Stage 0/I and
Stage II sources \citep{Indebetouw07}. However, these selection biases
become less important for more luminous sources, and the large number
of Stage 0/I objects with $M > 5~M_{\odot}$ found in our sample, 69, is
unprecedented among studies of Galactic star formation regions to date. It
implies that W51 is not only very massive but that it could also be
exceptionally young, even when compared to M17. This result is consistent
with the idea that W51 IRS 2E, the most luminous embedded source in
W51A, is a highly embedded, very young O star that has yet to produce
a detectable hypercompact H II region \citep{Figueredo08}.\footnote{
IRS 2E is saturated in the GLIMPSE and MIPSGAL images, hence it is not
included in our sample of candidate YSOs.}

\subsection{Color-Color Diagrams}

Figure \ref{ccd_yso_av10} shows the distribution of the final 437 YSO
candidates in IRAC color-color diagrams. The sources plotted in Figure
\ref{ccd_yso_av10} are classified according to their most probable
evolutionary Stage from the \cite{Robitaille06} models. Red dots are Stage
0/I YSOs that are relatively unevolved, heavily embedded in their natal
envelopes. Yellow dots are Stage II YSOs, like Class II T Tauri stars,
with optically thick circumstellar disks. There are no Stage III YSOs
with optically thin remnant disks in this sample. Green dots are objects
with ambiguous stage classification. In Figure \ref{ccd_yso_av10}({\it
a}), we see that the color selection criteria of \cite{Simon07}
includes many reddened stars as well as YSO candidates within the
solid line. If the color selection criteria are shifted by $\sim$
0.3 mag in the $[3.6]-[4.5]$ color (corresponding to $A_V = 20$ mag),
relatively clean YSO candidates are selected inside the dashed line. In
Figure \ref{ccd_yso_av10}({\it b}), if the ``disk domain'' defined by
\cite{Allen04} also moves vertically in the $[3.6]-[4.5]$ color, many
Stage II YSO candidates with optically thick circumstellar disks fall
in the disk domain region. The shift of the color selection criteria
in the direction of the reddening vector is consistent with the fact
that the W51 complex is a more massive star forming region with more
extinction than the star forming region studied by \cite{Allen04}. Figure
\ref{ccd_yso_av10}({\it c}) shows the importance of the 24 $\micron$
flux in classifying the evolutionary stages of YSO candidates. Stage
0/I objects (red dots) are redder than Stage II objects (yellow dots)
in $[8.0]-[24]$ color space. Most of the ambiguous sources (green dots)
are in the boundary region between Stage 0/I and Stage II.

%---------------------------------------------------------------------------
\subsection{Spatial Distribution of YSOs}

%------------------------------
\input{fig06.tex}

%------------------------------

Figure \ref{w51_stage_plot} shows the spatial distribution, classified
by evolutionary stage, of the 437 YSO candidates associated with the W51
region on the IRAC 8 \micron\ image. YSO candidates are marked in red
for Stage 0/I, yellow for Stage II, and green for ambiguous sources by
evolutionary stage. Stage 0/I YSOs are clearly concentrated toward the
bright \ion{H}{2} regions. Stage II YSOs are distributed outside of the
central dense region. Most sources within the bright 8 \micron\ emission
near the central region show strong clustering. These clustered sources
are associated with the W51 molecular cloud. Several clusters of YSOs are
also apparent in the outer region. A very strong concentration of YSOs
is found in an IR dark cloud (IRDC) centered on ($l,b$)=(49.4\arcdeg,
0.0\arcdeg). The detailed analysis of the CO observations and IRDC
associated with this YSO clustering will be presented elsewhere (Kang
et al., in preparation).

%------------------------------
\input{fig07.tex}

%------------------------------

Figure \ref{w51_mass_plot} shows the spatial distribution of the YSOs
classified by the mass of the central star on the IRAC 8 \micron\
image. The vicinity of the compact radio continuum sources marked as
crosses in Figure \ref{All_W51} is bright in the IRAC 8 \micron\ band
due to PAH emission features stimulated by UV radiation from \ion{H}{2}
regions. The mass range of YSOs is from 1 to 18 $M_\odot$. We divide all
YSOs into three mass ranges: $M_\star \leq 5 M_{\odot}$, $5 M_{\odot}
\leq M_\star < 8 M_{\odot}$, $M_\star \geq 8M_{\odot}$. Most of the
massive YSOs ($\geq 8 M_\odot$) are located near the main \ion{H}{2}
regions of W51. Clustering of massive YSOs is clearly apparent in the
central region, while the YSO cluster centered on ($l,b$)=(49.4\arcdeg,
0.0\arcdeg) consists of many lower mass YSOs and only one massive YSO. The
mass distribution of YSOs shows evidence of mass segregation with the most
massive objects exhibiting a strong concentration close to the center
of the \ion{H}{2} regions. We can quantify this trend by computing the
mass function of YSOs for various regions within our target area.

\subsection{Mass Function of YSOs}

We investigate the variation with spatial distribution of the mass
function (MF) of YSOs. We divide our target field ($1.0 \arcdeg \times
1.25 \arcdeg$) into two regions: the central part including very bright
emission in all bands from near-IR to 21 cm continuum ($48.80 \arcdeg
\le l \le 49.65 \arcdeg$ and $ -0.48 \arcdeg \le b \le -0.15 \arcdeg$
shown by the solid box in Figure \ref{w51_mass_plot}) and the outside
region excluding the central region. We also find the MF of the W51A
and W51B regions of the central part, separately.

%------------------------------
\input{fig08.tex}

%------------------------------

To obtain as clean a sample as possible, we remove from the final
source list those objects with a probability of being extragalactic
and AGB contaminants, as discussed in Section \ref{sec:est_cont}. To
compare MFs without bias of source selection we set two criteria. The
sensitivity limit of IRAC sources is affected by the brightness of the
diffuse emission, in particular at IRAC 8.0 $\micron$. Therefore, the
sensitivity limit in star-forming regions with bright diffuse emission
is certainly different from the outside region. \cite{Robitaille08}
found that sources with lower limits on the IRAC 4.5 and 8.0 \micron\
fluxes of 0.5 and 10 mJy respectively are not affected by the variations
in the diffuse emission on a Galaxy-wide scale. We apply these lower
limits on the IRAC 4.5 and 8.0 \micron\ fluxes to our final YSO sources,
which selects 113 YSOs. The mass distribution of 113 YSOs turns over
at 5 $M_\odot$. Therefore we use only those 94 YSOs with $M_\star > 5 ~
M_\odot$ for comparing the shape of the mass function of YSOs in the W51
region to avoid introducing bias due to sensitivity and completeness.
Although we have used very conservative brightness cutoffs in constructing
our mass functions, the mass functions are unlikely to be 100\% complete
in the central regions, due to the selection effects imposed by confusion
and extremely bright nebular emission. In the case of more luminous
objects, the stellar luminosity dominates in the total luminosity of the
source, while the accretion luminosity is clearly negligible, so the
SED-determined luminosity simply sets the corresponding stellar mass
given the distance. We find that the mass-luminosity relation for the
sources used in the MF derivation agrees with a recent re-evaluation
of the ZAMS mass-luminosity relation by \cite{Malkov07}, as shown in
Figure \ref{ML_relation}. The mean distance of 94 YSOs derived from
the SED fitter is $6.6\pm0.2$ kpc which is close to the $6.1 \pm 1.3$
kpc measured for water masers in the W51A region \citep{Imai02}.

The mass distribution from 94 YSOs which satisfy the sensitivity and
completeness criteria in our target region has a best-fit slope of the MF,
$\Gamma = -1.70 \pm 0.10$, as defined by 
\begin{displaymath}
M_\star^{\Gamma} \propto \frac{d \xi (M_\star)}{d (\mathrm{log}
M_\star)}
\end{displaymath}
for $M_\star > 5 M_\odot$ (solid line in Figure \ref{IMF_spatial}({\it
a})). This value is very close to the stellar IMF slope of $\Gamma =
-1.8$, derived by \cite{Rana87} for the solar neighborhood field stars
($1.5 M_\odot - 100 M_\odot$). Foreground sources which satisfy the same
sensitivity and completeness have the MF, $\Gamma = -2.13 \pm 0.29$
(dash-dotted line in Figure \ref{IMF_spatial}({\it a})). However the
YSO MFs of the central active star forming region ($\Gamma = -1.26 \pm
0.12$) and the outer region ($\Gamma = -2.36 \pm 0.26$) are significantly
different in Figure \ref{IMF_spatial}({\it b}). The MF for the inner
region has a slope similar to the Salpeter law, while the MF of the outer
part is very steep. \cite{Kroupa03} note that field-star IMFs are always
steeper for masses over $1 M_\odot$ than the stellar IMF derived from
an individual massive star cluster. In the Large and Small Magellanic
Clouds, \cite{Massey02} found a very steep IMF slope ($\Gamma \sim -4
\pm 0.5$) outside of the OB associations. The IMF is often observed to
be shallower in the central regions of clusters. This mass segregation
effect might occur because the most massive stars are either born near the
center or migrate toward the center. The question is then whether this
segregation is due to initial conditions or to dynamical evolutionary
effects. In the W51 \ion{H}{2} region complex, most of the massive YSOs
in the early stages of evolution are concentrated in the central region
(see Figure \ref{w51_stage_plot} and \ref{w51_mass_plot}). Because
we derive the MFs using YSOs in the early stages, still surrounded by
disks and/or infalling envelopes, these YSOs must be located in their
birthplaces. Therefore the spatial variation of the MF in W51 indicates
that the mass segregation is caused by initial conditions, not dynamical
effects. The difference between the central and outer regions is whether
the sources are associated with the most active star-forming region
or not. In the outer region YSOs may be forming in relatively small or
isolated molecular clouds.

%------------------------------
\input{fig09.tex}

%------------------------------

We compare MF of YSOs calculated from {\it Spitzer} observation to
the stellar mass function. IMFs of many dense clusters in the solar
neighborhood are close to the Salpeter IMF slope, $\Gamma = - 1.35$
on a plot with log-mass intervals. \cite{Scalo86, Scalo98} reviewed
the determination of the stellar IMF extensively and concluded the
IMF slope of field stars to be $-1.7\pm0.5$ in the range of $2-10 \,
M_\odot$. \cite{Ninkov95} found a slope of $\Gamma = -1.38 \pm 0.19$
for masses between $2.5-30 M_\odot$ from the OB cluster in the \ion{H}{2}
region IC 1805 in the Cas OB6 association associated with the molecular
cloud W4. They noted that the mass function determined from young
galactic OB clusters varies from $\Gamma=-1.0$ to $-1.4$. \cite{Garmany82}
determined a slope of the IMF of $\Gamma=-1.6$ for massive stars ($> 20
M_\odot$) within 2 kpc of the Sun. They found evidence that the slope of
the IMF varied with galactic radius: the IMF of stars inside the solar
circle is $\Gamma=-1.3$ and outside is $\Gamma=-2.1$. They suggested
that the excess of massive stars inside the solar circle was due to the
OB associations in the Carina and Cygnus spiral arms. \cite{Hunter96}
estimated an IMF slope of $-1.6 \pm 0.7$ for intermediate-mass stars
($6.5-18 M_\odot$) from NGC 604 in the galaxy M33. \cite{Massey98}
derived a slope close to the Salpeter value from the 30 Doradus region
of the LMC, suggesting that there is no strong dependence of the IMF
slope on metallicity. \cite{Sabbi08} found a present-day mass function
of $\Gamma = -1.43 \pm 0.18$ of NGC 346, the active star-forming region
in the SMC. \cite{Homeier05} derived a slope of $\Gamma = -1.6 \pm 0.3$
in the giant radio \ion{H}{2} region W49A. \cite{Okumura00} found
that the IMF of the massive stars ($M\geq 4 M_\odot$) in G49.5-0.4
of W51A is consistent with an IMF slope of $-1.8$, but that the IMF
shows a statistically significant excess of stars in the high mass
range above $30 M_\odot$. These various observations suggest that low
density regions have slightly steeper IMFs than the Salpeter slope,
while more massive clusters have a more massive upper end of the IMF
\citep{Elmegreen04,Elmegreen06}. The results we present here appear to
support this conclusion.

We divide the central active star-forming region into W51A and W51B
(Figure \ref{w51_mass_plot}) and find the slopes of the MF of $\Gamma
= -1.17 \pm 0.26$ and $\Gamma = -1.32 \pm 0.26$, respectively (Figure
\ref{IMF_spatial}({\it c})). The slopes are the same as the Salpeter IMF
within the errors. Similar results that the MFs of YSOs are consistent
with the Salpeter IMF are reported from other active star-forming regions,
e.g., M17 in the Galaxy \citep{Povich09}, N66 in the SMC \citep{Simon07},
and the LMC \citep{Whitney08}. It is not surprising that our MF of
YSOs in this active star-forming region is similar to the stellar IMF
because even the core mass function, which has $\Gamma = -1.3 $ for $M
> 0.8 M_\odot$ in Perseus, Serpens, and Ophiuchus, is consistent with
the Salpeter IMF \citep{Enoch08}. In a future paper, we will present
the detailed properties of molecular clouds associated with the YSOs in
W51A and W51B using our \co and \tco $J=2-1$ emission line maps of the
region studied in the present work.

%%%%%%%%%%%%%%%%%%%%%%%%%%%%%%%%%%%%%%%%%%%%%%%%%%%%%%%%%%%%%%%%%%%%%%%%%%%
\section{Summary}
\label{sec:summary}

We have found 737 candidate YSOs in a $1.25 \arcdeg \times 1.00\arcdeg
$ area near the W51 GMC using {\it Spitzer Space Telescope} data. We
distinguish YSOs from main-sequence or red giant stars, asymptotic giant
branch stars, and background galaxies by fitting model SEDs to fluxes
of sources. We divide the total of 737 YSOs into two groups based on
interstellar extinction for each YSO: sources associated with the W51
region and foreground sources. We identify 69 highly embedded Stage 0/I
candidate YSOs in our field with masses ${>}~5 ~M_{\odot}$ (corresponding
to mid- to early-B main-sequence spectral types) 46 of which are located
in the central active star forming regions of W51A and W51B. From the
YSOs associated with W51, we find evidence for mass segregation showing
that the most massive YSOs are concentrated on the W51 \ion{H}{2}
region complex. We find a variation in the spatial distribution of the
MF of YSOs in the mass range between 5 $M_\odot$ and 18 $M_\odot$. The
mass distribution from YSOs in our entire target region has a slope of
$\Gamma = -1.70 \pm 0.10$. When the sample is divided into the active
star-forming region and the surrounding outer region, the slope of MFs
in the star-forming region ($\Gamma =-1.26\pm0.12$) is shallower than
that of outer region ($\Gamma=-2.36\pm0.26$). However, we do not find
any difference in slopes between two active star forming regions within
the complex: $\Gamma = -1.17 \pm 0.26$ and $\Gamma = -1.32 \pm 0.26$ in
W51A and W51B, respectively. The variation of the MF for YSOs embedded
in the molecular cloud implies that the distribution of stellar masses
in clusters depends on the local conditions in the parent molecular cloud.

\acknowledgments

We thank Barbara A. Whitney for a helpful discussion in the very early
stage of this work. We thank Brian L. Babler and Marilyn R. Meade for
help with MIPS photometry and Bon-Chul Koo for providing 21 cm radio
continuum data. The authors also thank the anonymous referee for helpful
comments that improved the text. M.\,K. also thanks Yujin Yang and Amy
Stutz for helpful discussion. This research was supported in part by NSF
grant AST-0708131 to the University of Arizona. This work was supported
by the Korea Research Foundation Grant funded by the Korean Government
(MOEHRD: KRF-2007-612C00050). MSP is supported by an NSF Astronomy and
Astrophysics Postdoctoral Fellowship under award AST-0901646.

%%%%%%%%%%%%%%%%%%%%%%%%%%%%%%%%%%%%%%%%%%%%%%%%%%%%%%%%%%%%%%%%%%%%%%%%%%%
% References
%%%%%%%%%%%%%%%%%%%%%%%%%%%%%%%%%%%%%%%%%%%%%%%%%%%%%%%%%%%%%%%%%%%%%%%%%%%

\end{document}

%% file: fig01.tex
\begin{figure}
\plotone{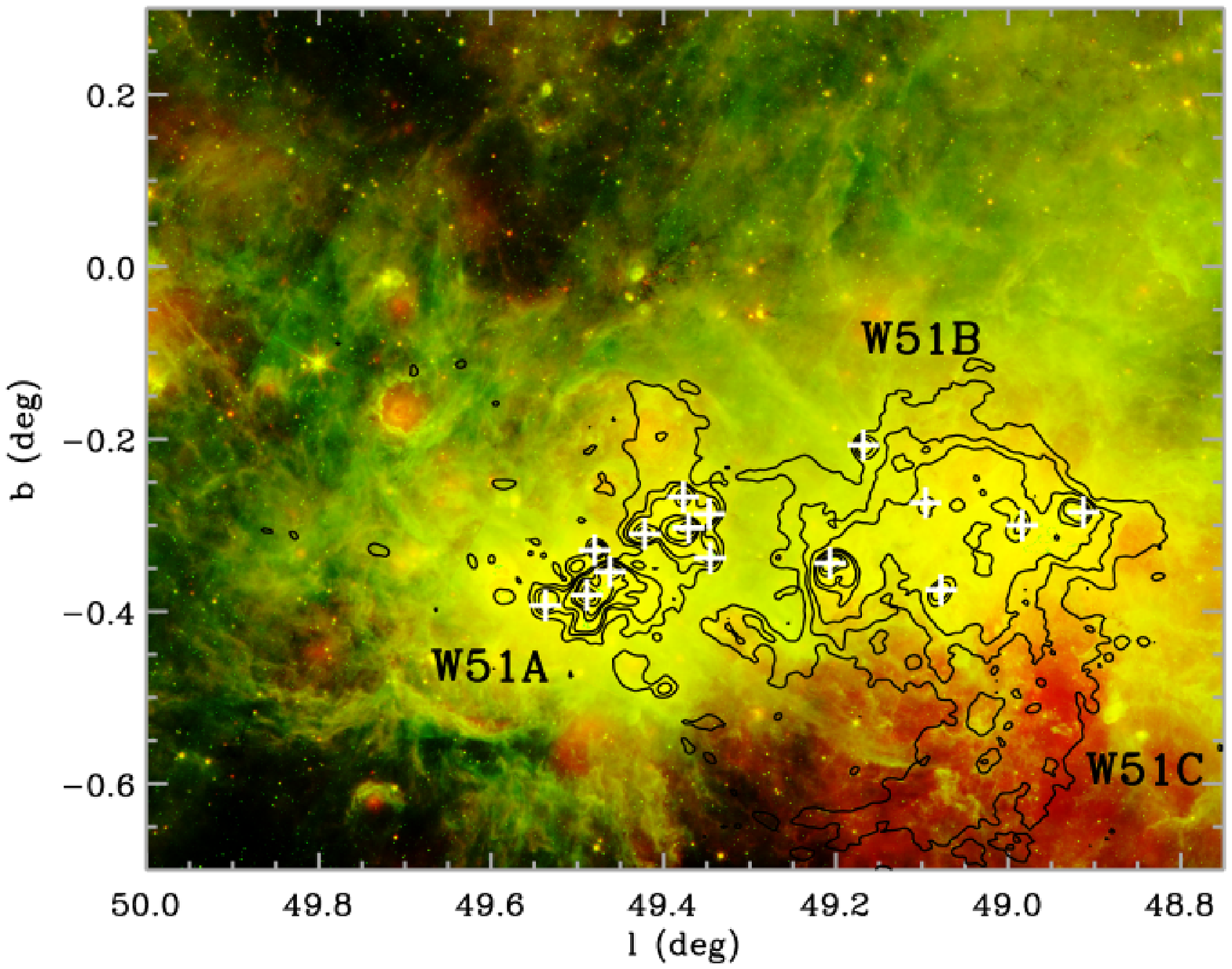}

\caption{ A color image of the W51 complex composed of MIPS 24
$\micron$
(red) and IRAC 8.0 $\micron$ (green). Hot dust grains and PAH
dominated
emission appear in red and green, respectively. The contours show 21
cm
radio continuum emission made by combining VLA data and Effelsberg
100-m
telescope data \citep{Koo97}. The contour levels are $-$0.015, 0.015,
0.05, 0.10, 0.3, 0.5, 1.0, 1.5, and 2.4 Jy beam$^{-1}$. Cross symbols
mark the radio continuum sources listed by \cite{Koo97}.}

\label{All_W51}
\end{figure}

%% file: fig02.tex
\begin{figure*}
%\epsscale{0.8}
\plotone{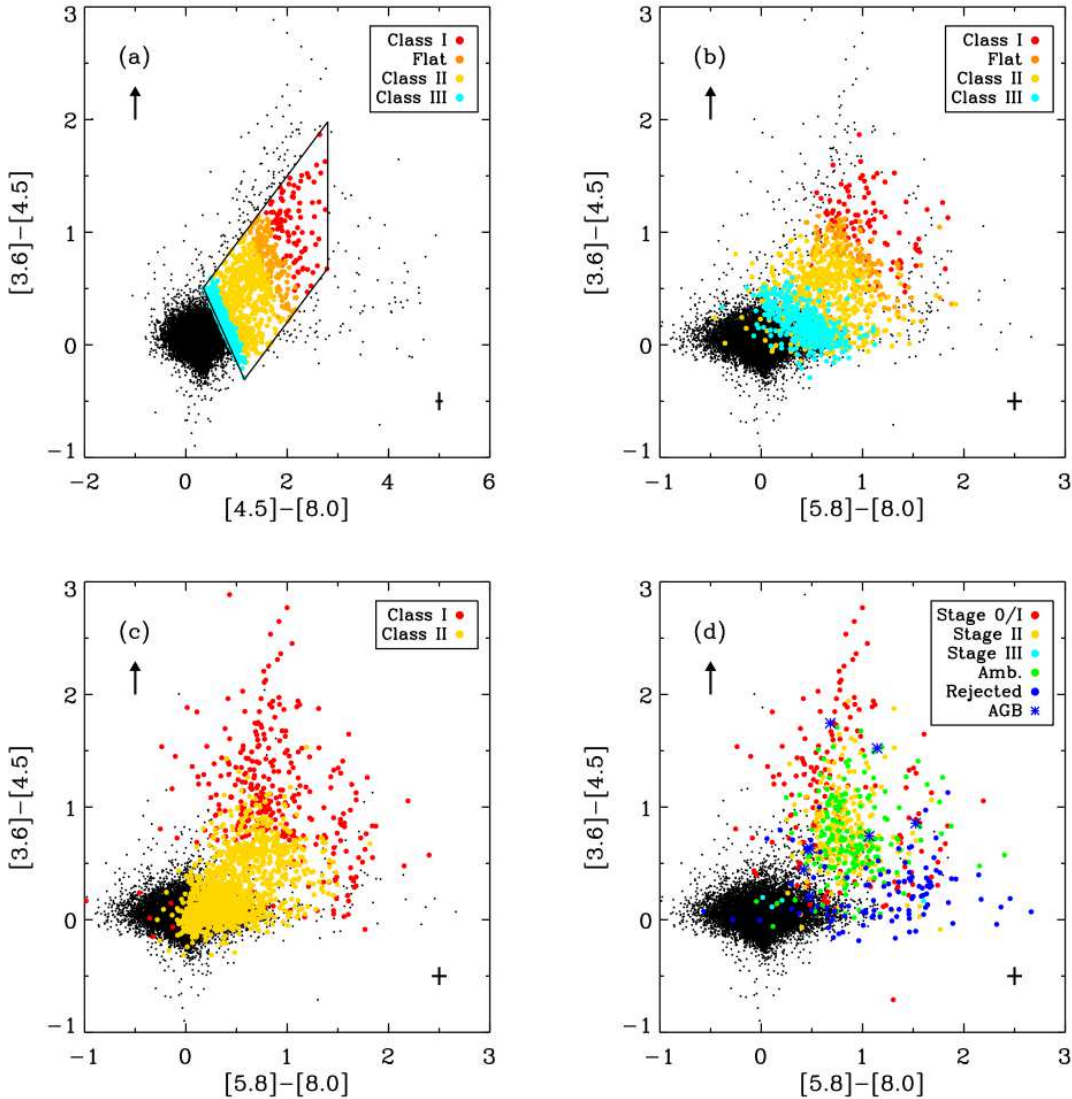}

\caption{({\it a}) IRAC $[3.6]-[4.5]$ vs. $[4.5]-[8.0]$ color-color
diagram by the color selection criteria of \cite{Simon07}. All sources
were detected in all 4 IRAC bands in the GLIMPSE Catalog. YSO candidates
classified by the spectral index $\alpha$ are marked in red for Class I
objects, orange for the flat-spectrum sources, yellow for the Class II
sources, and cyan for the Class III sources. The solid line represents the
color selection criteria of \cite{Simon07}. ({\it b}) IRAC $[3.6]-[4.5]$
vs. $[5.8]-[8.0]$ color-color diagram by the color selection criteria of
\cite{Simon07}. ({\it c}) IRAC $[3.6]-[4.5]$ vs. $[5.8]-[8.0]$ color-color
diagram by the color selection criteria of \cite{Gutermuth08}. Class
I objects are marked in red and Class II objects are marked in yellow.
({\it d}) IRAC $[3.6]-[4.5]$ vs. $[5.8]-[8.0]$ color-color diagram with
YSO candidates identified using the SED fitting tool. Small black dots
are sources well-fit by stellar photosphere SEDs. YSO candidates are
marked in red for Stage 0/I, yellow for Stage II, cyan for Stage III, and
green for ambiguous sources. AGB candidates are marked by blue asterisk
symbols. Blue dots are well-fit YSO candidates that were discarded
from the final sample. Reddening vector for A$_{V}$=20 mag based on
the extinction laws of \cite{Indebetouw05} is shown as a filled arrow.
Black crosses show typical photometric errors.}

\label{ccd_f2}
\end{figure*}

%% file: table01.tex
\begin{deluxetable}{lr}
\tabletypesize{\scriptsize}
\tablecaption{Source Counts in the W51 Region\tablenotemark{a}}
\tablewidth{0pt}
\tablehead{{Sources}&{Number}}
\startdata

 In GLIMPSE Catalog                                                &
104,582 \\
 Fit with SED models ($N_{data} \geq 4$)                           &
70,767 \\
 Well-fit by stellar photosphere SEDs ($\chi^{2}/N_{data} \leq 4$) &
69,316 \\
 Poorly-fit by stellar photosphere SEDs ($\chi^{2}/N_{data} > 4$)  &
1,451 \\
 Well-fit by YSO SEDs ($\chi^{2}/N_{data} \leq 4$)                 &
1,327 \\
 In final sample of YSO candidates                                 &
737 

\enddata
\tablenotetext{a}{ The 1.25\arcdeg $\times$ 1.0\arcdeg region centered
on (l,b) = (49.375\arcdeg, $-$0.2\arcdeg)}

\label{table1}

\end{deluxetable}

%% file: table02.tex
\begin{deluxetable*}{rcrrrrrrrrr}[!b]
\tabletypesize{\scriptsize}
\tablecaption{Magnitudes of YSO Candidates}
\tablewidth{0pt}
\tablehead{
\colhead{No.} & \colhead{IRAC designation} &
\multicolumn{8}{c}{Magnitudes} & {Assoc.}\tablenotemark{a} \\
\cline{3-10}
&&
\colhead{$J$} &
\colhead{$H$} &
\colhead{${K_s}$} &
\colhead{$[3.6]$} &
\colhead{$[4.5]$} &
\colhead{$[5.8]$} &
\colhead{$[8.0]$} &
\colhead{$[24]$}  &
 }
\startdata

  1&  SSTGLMC G048.7567$-$00.6341&     15.89&     14.81&     14.00&
12.58&     12.15&     11.85&     10.94&      7.43&     F\\
  2&  SSTGLMC G048.7579$-$00.2797&     14.17&     13.37&     12.94&
12.40&     12.30&     11.66&   \nodata&      6.03&     F\\
  3&  SSTGLMC G048.7605$-$00.0388&   \nodata&   \nodata&   \nodata&
13.05&     12.28&     11.07&     10.02&      6.24&     W\\
  4&  SSTGLMC G048.7618$+$00.0627&   \nodata&   \nodata&   \nodata&
14.33&     13.20&     11.68&     10.94&      7.62&     W\\
  5&  SSTGLMC G048.7637$+$00.2022&   \nodata&   \nodata&   \nodata&
13.80&     12.37&     11.27&     10.67&      6.47&     W\\
  6&  SSTGLMC G048.7655$+$00.1017&     14.91&     14.08&     13.41&
12.74&     12.64&   \nodata&   \nodata&      8.44&     F\\
  7&  SSTGLMC G048.7667$+$00.0766&     15.97&     14.10&     12.82&
11.33&     10.80&     10.31&      9.84&      7.60&     F\\
  8&  SSTGLMC G048.7702$-$00.1505&   \nodata&   \nodata&   \nodata&
13.58&     12.15&     11.46&     11.06&      6.26&     W\\
  9&  SSTGLMC G048.7703$+$00.0786&   \nodata&   \nodata&   \nodata&
12.41&     10.88&      9.67&      8.85&      5.81&     W\\
 10&  SSTGLMC G048.7720$-$00.4792&     16.27&     14.38&     12.90&
10.94&     10.53&     10.03&      9.25&   \nodata&     F\\
 11&  SSTGLMC G048.7726$-$00.2215&   \nodata&   \nodata&     13.20&
12.06&     11.74&      9.94&   \nodata&      2.28&     F\\
 12&  SSTGLMC G048.7744$+$00.1091&   \nodata&   \nodata&     14.87&
11.70&     10.47&      9.72&      9.13&      7.20&     W\\
 13&  SSTGLMC G048.7750$-$00.1507&   \nodata&   \nodata&   \nodata&
13.22&     12.08&     10.41&   \nodata&      5.34&     W\\
 14&  SSTGLMC G048.7772$+$00.2513&   \nodata&     14.94&     13.88&
12.92&     13.63&     11.11&      9.80&      3.72&     W\\
 15&  SSTGLMC G048.7780$+$00.2275&   \nodata&     15.19&     14.29&
13.20&     12.72&     12.15&     11.56&   \nodata&     F

\enddata
\tablenotetext{a}{ Association by the foreground extinction $A_V$ - W,
W51 source; F, foreground source}
\tablecomments{Table 2 is published in its entirety in the electronic
edition
of the Astrophysical Journal.
A portion is shown here for guidance regarding its form and content.}

\label{table:mag}
\end{deluxetable*}

%% file: table03.tex
\begin{deluxetable*}{rcrrrrrrrrrrrr}

\tabletypesize{\scriptsize}
\tablecaption{Model parameters for YSO candidates}
\tablewidth{0pt}
\tablehead{\colhead{No.} &
           \colhead{Name} &
           \multicolumn{3}{c}{$A_V$ (mag)} &
           \multicolumn{3}{c}{$M_\star$ ($M_\odot$)} &
           \multicolumn{3}{c}{$L_{tot}$ ($L_\odot$)} &
           \colhead{Evol.} & {Class\tablenotemark{c}} &
{Class\tablenotemark{d}}\\
           \cline{3-5} \cline{6-8} \cline{9-11}
            &
           \colhead{(G$l+b$)}&
           \colhead{L\tablenotemark{a}} &
\colhead{Ave\tablenotemark{b}} & \colhead{U\tablenotemark{a}} &
           \colhead{L\tablenotemark{a}} &
\colhead{Ave\tablenotemark{b}} & \colhead{U\tablenotemark{a}} &
           \colhead{L\tablenotemark{a}} &
\colhead{Ave\tablenotemark{b}} & \colhead{U\tablenotemark{a}} &
           \colhead{Stage} & {$\alpha_{IRAC}$} & {$\alpha_{2-24}$}}
\startdata

  1&  G048.7567$-$00.6341&     2.9&     4.0                  &
5.4&    2.6&     3.0                 &     3.1&     43&       74
&       86&    II &       II&    II\\
  2&  G048.7579$-$00.2797&     0.3&     1.6                  &
2.6&    1.4&     2.7                 &     4.0&     17&       33
&       45&   0/I &  \nodata&     F\\
  3&  G048.7605$-$00.0388&    23.1&    33.3                  &
42.8&    3.3&     4.1                 &     4.8&     87&      269
&      403&    II &        I&     I\\
  4&  G048.7618$+$00.0627&    43.5&    51.8                  &
60.0&    3.3&     4.0                 &     4.7&    122&      274
&      383&    II &        I&     F\\
  5&  G048.7637$+$00.2022&     6.6&    25.7                  &
53.8&    2.2&     3.5                 &     4.7&     49&      155
&      311&   0/I &        I&     I\\
  6&  G048.7655$+$00.1017&     1.6&     3.0                  &
4.3&    2.8&     3.2                 &     3.8&     16&       66
&      122&   Amb &  \nodata&    II\\
  7&  G048.7667$+$00.0766&     7.1&     7.5                  &
8.1&    3.3&     4.0                 &     4.5&    129&      229
&      327&    II &       II&    II\\
  8&  G048.7702$-$00.1505&     1.2&    18.8                  &
39.6&    1.8&     3.9                 &     5.7&     35&      106
&      188&   0/I &        F&     I\\
  9&  G048.7703$+$00.0786&    46.5&    52.7                  &
60.0&    5.4&     6.6                 &     7.8&    658&     1788
&     2668&    II &        I&     F\\
 10&  G048.7720$-$00.4792&     6.2&     9.3                  &
11.4&    3.8&     4.6                 &     5.5&    175&      349
&      456&    II &       II&     F\\
 11&  G048.7726$-$00.2215&     1.1&     7.9                  &
14.7&    4.8&     6.0                 &     7.2&    190&      472
&      753&   0/I &  \nodata&     I\\
 12&  G048.7744$+$00.1091&    24.0&    34.7\tablenotemark{*} &
40.1&    5.4&     5.8\tablenotemark{*}&     6.0&   1179&     1464
&     1536&    II &        F&     F\\
 13&  G048.7750$-$00.1507&     4.9&    32.2                  &
56.8&    1.3&     4.3                 &     6.0&     46&      663
&      809&    II &  \nodata&     I\\
 14&  G048.7772$+$00.2513&    13.7&    36.8                  &
58.9&    2.0&     3.6                 &     5.2&     98&      238
&      389&   0/I &        I&     I\\
 15&  G048.7780$+$00.2275&     3.7&     8.2                  &
11.8&    1.4&     2.6                 &     3.7&     15&       35
&       43&   Amb &       II&    II

\enddata
\tablenotetext{a}{The range of values reported here on the
extinctions,
masses, and luminosities for each parameter are the lower and upper
limit of 68\% probability for all of the acceptable YSO models.}
\tablenotetext{b}{Asterisks indicate ranges of 95\% probability.}
\tablenotetext{c}{YSO class based on spectral slope $\alpha$ using the
flux densities detected in all four IRAC bands.}
\tablenotetext{d}{YSO class based on spectral slope $\alpha$ using the
flux densities available between 2 and 24 \micron.}
\tablecomments{Table 3 is published in its entirety in the
electronic edition of the Astrophysical Journal. A portion is shown
here
for guidance regarding its form and content.}
\label{table:parameters}

\end{deluxetable*}

%% file: table04.tex
\begin{deluxetable*}{lrrrrrrrrrr}
\tabletypesize{\scriptsize}
\tablecaption{Number of YSO candidates for Stages and Classes}
\tablewidth{0pt}
\tablehead{\colhead{Region} &
           \colhead{No.} &
           \multicolumn{4}{c}{Physical} &&
           \multicolumn{4}{c}{Observational} \\
           \cline{3-6} \cline{8-11}
           & &
           \colhead{Stage 0/I} & {II}   & {III} & {Amb.} &&
           \colhead{Class I}   & {Flat} & {II}  & {III}
           }
\startdata
\multicolumn{11}{c}{Sources identified as YSOs with the SED fitter}\\
\hline
All  \tablenotemark{a}& 737& 228& 255& 5& 249&&  256& 243& 235& 3\\
Fore.\tablenotemark{b}& 300& 100&  93& 5& 102&&   57&  75& 165& 3\\
W51  \tablenotemark{c}& 437& 128& 162& 0& 147&&  199& 168&  70& 0\\
Outer\tablenotemark{d}& 281&  64& 137& 0&  80&&   88& 132&  61& 0\\
W51A \tablenotemark{e}&  72&  31&   9& 0&  32&&   52&  16&   4& 0\\
W51B \tablenotemark{f}&  84&  33&  16& 0&  35&&   59&  20&   5& 0\\
\hline
\multicolumn{11}{c}{Sources detected in all four IRAC bands and MIPS
24 \micron }\\
\hline
All                   & 305&  61& 196& 4&  44&&   85& 108& 112& 0\\
Fore.                 & 109&  26&  61& 4&  18&&   17&  23&  69& 0\\
W51                   & 196&  35& 135& 0&  26&&   68&  85&  43& 0\\
Outer                 & 164&  24& 118& 0&  22&&   45&  78&  41& 0\\
W51A                  &  15&   8&   6& 0&   1&&   12&   2&   1& 0\\
W51B                  &  17&   3&  11& 0&   3&&   11&   5&   1& 0\\
\hline
\multicolumn{11}{c}{Sources classified as YSOc by \cite{Harvey07}
criteria}\\
\hline
All                   & 270&  59& 169& 1&  41&&   83&  97&  90& 0\\
Fore.                 & 101&  26&  57& 1&  17&&   17&  23&  61& 0\\
W51                   & 169&  33& 112& 0&  24&&   66&  74&  29& 0\\
Outer                 & 137&  22&  95& 0&  20&&   43&  67&  27& 0\\
W51A                  &  15&   8&   6& 0&   1&&   12&   2&   1& 0\\
W51B                  &  17&   3&  11& 0&   3&&   11&   5&   1& 0
\enddata
\tablenotetext{a}{Entire sample of YSOs identified by SED fitter}
\tablenotetext{b}{Foreground YSOs with $A_V < 10$ mag}
\tablenotetext{c}{W51 YSOs with $A_V \ge 10$ mag}
\tablenotetext{d}{YSOs with $A_V \ge 10$ mag and located in the
outside region
excluding the central region  ($48.80 \arcdeg \le l \le 49.65 \arcdeg$
and $ -0.48 \arcdeg \le b \le -0.15 \arcdeg$)}
\tablenotetext{e}{YSOs with $A_V \ge 10$ mag around W51A  ($49.30
\arcdeg \le l \le 49.65
\arcdeg$ and $ -0.48 \arcdeg \le b \le -0.15 \arcdeg$)}
\tablenotetext{f}{YSOs with $A_V \ge 10$ mag around W51B  ($48.80
\arcdeg \le l \le 49.30
\arcdeg$ and $ -0.48 \arcdeg \le b \le -0.15 \arcdeg$)}
\label{table:stage_class}
\end{deluxetable*}

%% file: fig03.tex
\begin{figure}
\epsscale{1.}
\plotone{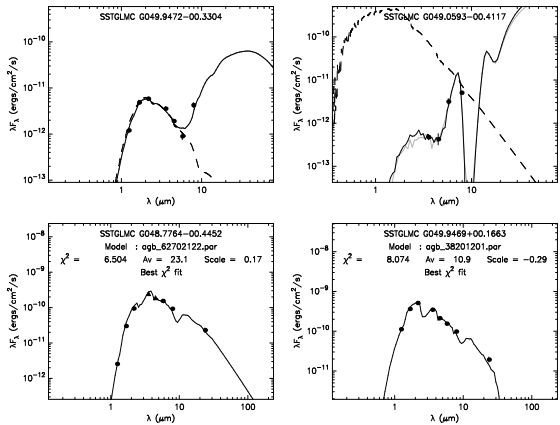}

\caption{Top: Example plots of rejected sources from final
samples. {\it Left}: Source with IR excess emission only in the IRAC
8.0 \micron\ band, but not detected in the MIPS 24 \micron\ band. {\it
Right}: Source with contamination by diffuse background emission in
5.8 \micron\ and 8.0 \micron\ bands. Bottom: Example SEDs of AGB
candidates. Triangles are datapoints with saturated fluxes that were
used as lower limits for the model fitting.}

\label{contamination}
\end{figure}

%% file: fig04.tex
\begin{figure}
\plotone{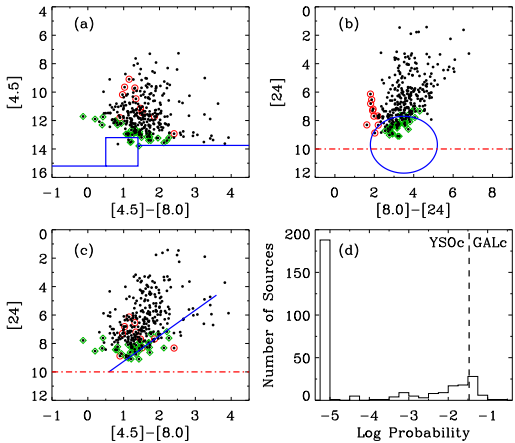}

\caption{({\it a}), ({\it b}), ({\it c}) Color-magnitude diagrams of
YSO candidates in W51. The solid lines show the color-magnitude cuts
for defining the YSO candidate criterion developed by \cite{Harvey07}
but with a total extinction of $A_V = 25$ mag along the line of sight,
rather than the average value of 7.5 mag for the Serpens cloud. The
dot-dashed lines show hard limits, fainter than which objects are
excluded
from the YSO category. Open symbols marked by circles and diamonds
are objects which may be AGBs and extragalactic sources. ({\it d})
Plot of the number of sources vs. probability of being a background
contaminant. The vertical dashed line shows the separation chosen by
\cite{Harvey07} for YSOs vs. extragalactic candidates.}

\label{harvey_cont}
\end{figure}

%% file: fig05.tex
\begin{figure*}
\plotone{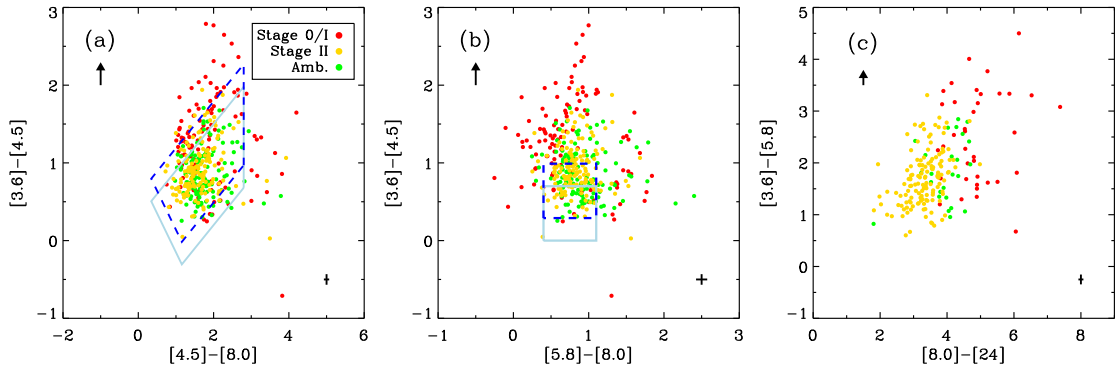}

\caption{({\it a}) IRAC $[3.6]-[4.5]$ vs. $[4.5]-[8.0]$ color-color
diagram with YSO candidates selected by the SED fitting tool. Points
in red mark Stage 0/I, yellow mark Stage II, and green mark ambiguous
sources. The solid line represents the region of the color selection
criteria of \cite{Simon07}. The dashed line shows the region shifted by
the reddening vector for the general visual extinction toward W51 of 20
mag \citep{Goldader94}. ({\it b}) IRAC $[3.6]-[4.5]$ vs. $[5.8]-[8.0]$
color-color diagram. The solid line represents the disk domain
of \cite{Allen04}. The dashed line shows the region shifted by the
reddening vector. ({\it c}) $[3.6]-[5.8]$ vs. $[8.0]-[24]$ color-color
diagram. Reddening vector for A$_{V}$=20 mag based on the extinction
laws of \cite{Indebetouw05} and Stutz et al. (2009, in preparation)
is shown as an arrow. Black crosses show typical photometric errors.}

\label{ccd_yso_av10}
\end{figure*}

%% file: fig06.tex
\begin{figure*}
\epsscale{0.85}
\plotone{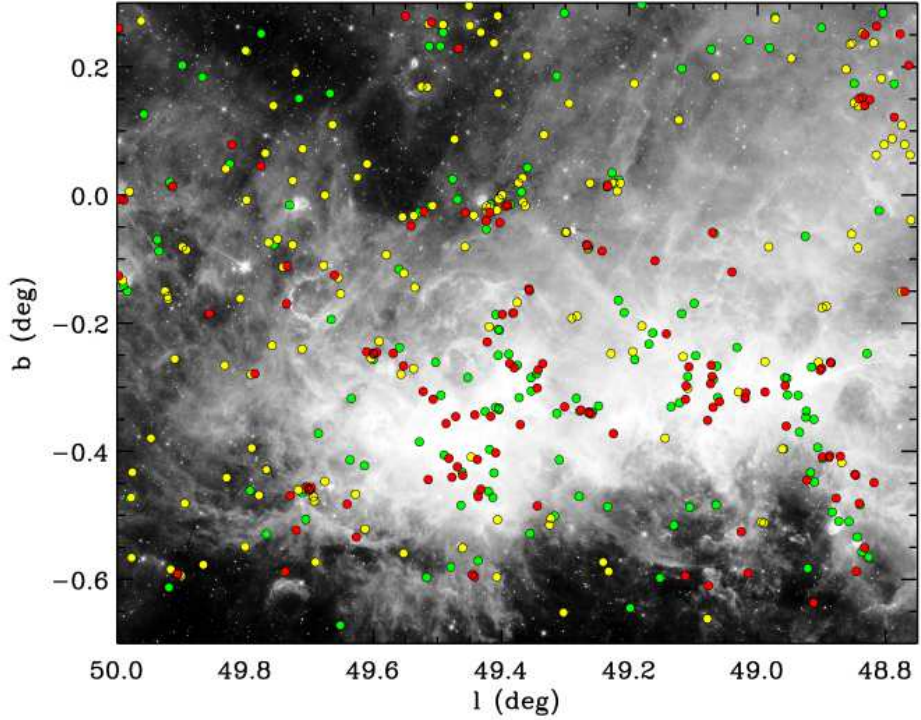}

\caption{Spatial distribution of YSOs in W51, overplotted on IRAC 8.0
$\micron$ image. YSO candidates are marked in red for Stage 0/I,
yellow for Stage II, and green for ambiguous sources.}

\label{w51_stage_plot}
\end{figure*}

%% file: fig07.tex
\begin{figure*}
\epsscale{0.85}
\plotone{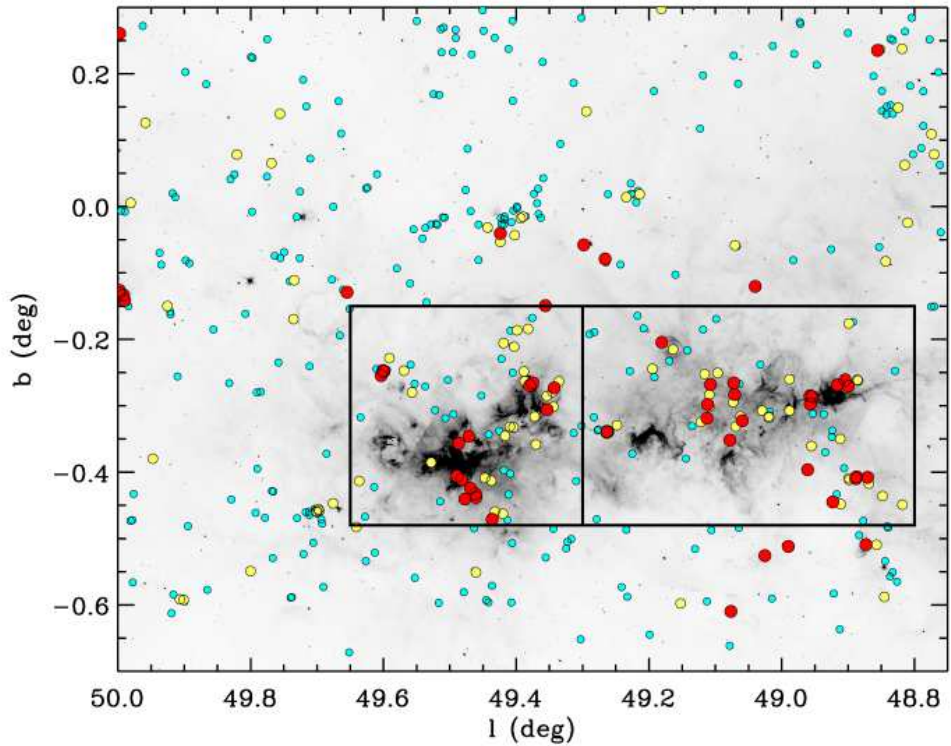}

\caption{The mass distribution of 437 YSOs associated with the W51 region
overplotted on the IRAC 8.0 $\micron$ image (This is the same image as
in Figure \ref{w51_stage_plot} but displayed with inverted grayscale
image to show more clearly the structure of very strong PAH emission
in black). Small (cyan), medium (yellow) and large (red) dots represent
YSOs with $M \leq 5 M_{\odot}$, $5 M_{\odot} \leq M < 8 M_{\odot}$ and
$M \geq 8M_{\odot}$. Box shows the boundary of the central region for
comparing the shape of the IMF. Left side of the box is the W51A region
and right side is the W51B region.}

\label{w51_mass_plot}
\end{figure*}

%% file: fig08.tex
\begin{figure}
\plotone{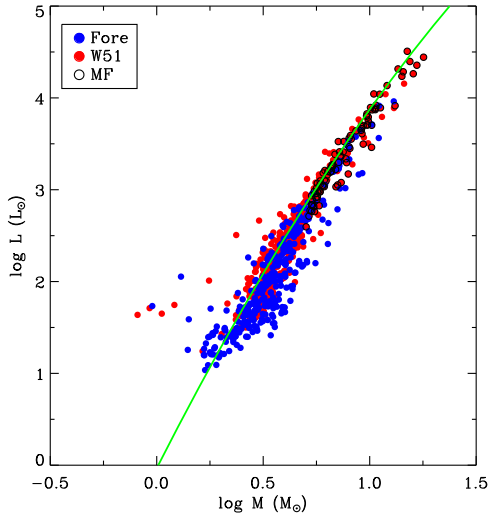}

\caption{The mass-luminosity relation for YSOs in the W51 region. Blue
and red dots represent foreground and W51 sources, respectively. Open
circles are the sources used for MF derivation. Green line shows the
ZAMS mass-luminosity relation by \cite{Malkov07} for intermediate mass
stars ($1 - 30 M_\odot$).}

\label{ML_relation}
\end{figure}

%% file: fig09.tex
\begin{figure*}
\plotone{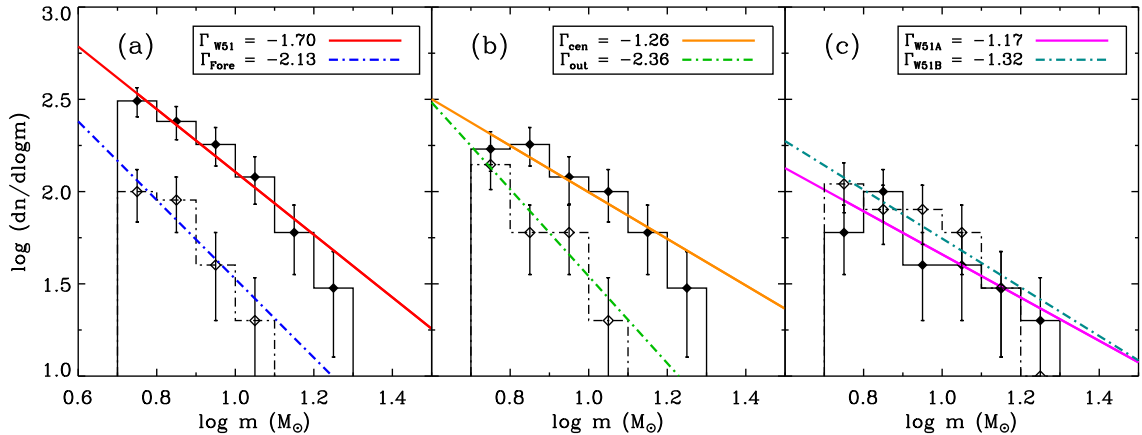}

\caption{({\it a}) The mass function for YSO candidates after
selecting only $>$ 0.5 and 10 mJy at 4.5 and 8.0 $\micron$ and cut off
$>$ 5 $M_\odot$. The solid line and the dash-dotted line represent the
mass function of YSOs associated with W51 and foreground YSOs,
respectively. ({\it b}) The mass function of the central region
(box in Figure \ref{w51_mass_plot}) (solid line) and the outside
region
(dash-dotted line). ({\it c}) The mass function of the W51A (solid
line)
and the W51B (dash-dotted line) region. All histograms are shown with
$\sqrt{N}$ errorbars.}

\label{IMF_spatial}
\end{figure*}